\documentclass[twocolumn,prb,preprintnumbers,superscriptaddress]{revtex4}
\usepackage{graphicx}
\usepackage{epstopdf}
\usepackage{amsmath}
\usepackage{amssymb} 
\usepackage{xcolor}
\usepackage[flushleft]{threeparttable}

\usepackage[normalem]{ulem}
\usepackage{hyperref}

\begin{document}
\title{1D transition metal oxide chains as a challenging model for ab initio calculations}

\author{Jila Amini}
\affiliation{ Department of Physics, Isfahan University of Technology, Isfahan 84156-83111, Iran }

\author{Mojtaba Alaei}
\email[ Correspondence email address: ]{m.alaei@iut.ac.ir}
\affiliation{ Department of Physics, Isfahan University of Technology, Isfahan 84156-83111, Iran }
\affiliation{ Skolkovo Institute of Science and Technology, 121205, Bolshoy Boulevard 30, bld. 1, Moscow, Russia }

\author{Stefano de Gironcoli}
\affiliation{ Scuola Internazionale Superiore di Studi Avanzati, Trieste, Italy }
\affiliation{ CNR-IOM DEMOCRITOS, Istituto Officina dei Materiali, Trieste, Italy }

\begin{abstract}
Providing highly simplified models of strongly correlated electronic systems that challenge ab initio calculations can serve as a valuable testing ground to improve these methods. In this study, we present a comprehensive investigation of the structural, magnetic, and electronic properties of one-dimensional transition metal mono-oxide chains (VO, CrO, MnO, FeO, CoO, and NiO) using density functional theory (DFT), DFT+$U$, and coupled-cluster singles and doubles (CCSD) calculations. The Hubbard $U$ parameter for DFT+$U$ is determined using linear response theory. In all systems studied except MnO, the presence of multiple local minima—primarily due to the electronic degrees of freedom associated with the d-orbitals—leads to significant challenges for DFT, DFT+$U$, and Hartree–Fock methods in finding the global minimum in ab initio calculations. 
Our results indicate that the antiferromagnetic (AFM) state is energetically favored for all chains, except CrO, when using DFT+$U$ and the Perdew–Burke–Ernzerhof (PBE) functional. Analysis of the band structures shows that while PBE often predicts metallic or half-metallic FM states, DFT+$U$ opens band gaps and correctly yields insulating behavior in all cases. Furthermore, we compared the energy differences between the AFM and FM states using DFT+$U$ and CCSD for CrO, MnO, FeO, CoO, and NiO. Our findings indicate that CCSD predicts larger energy differences in some cases compared to DFT+$U$, suggesting that the Hubbard $U$ parameter obtained through linear response theory may be overestimated when used to calculate energy differences between different magnetic states. For CrO, CCSD predicts an AFM ground state, in contrast to the predictions from DFT+$U$ and PBE methods.
\end{abstract}
\newcommand{\etal}{{\em et al}}
\maketitle

\section{INTRODUCTION}
One of the fundamental challenges in materials science is solving the Schr{\"o}dinger equation for many-body systems efficiently and accurately to predict material properties. Density functional theory (DFT) is among the most widely used methods for this purpose~\cite{burke2012perspective}. However, it has notable limitations, particularly when applied to systems with localized electrons~\cite{cohen2008insights}. A key issue is the self-interaction error, which can lead to inaccuracies in predicting electronic energy levels, such as band gaps and magnetic states. To mitigate this, the DFT+$U$ approach~\cite{DFTU1, DFTU2, DFTU3} introduces a correction term that improves the description of localized orbitals.

A major challenge in DFT+$U$ is determining the Hubbard \(U\) parameter accurately to ensure consistency with experimental results. Recently, we demonstrated that using linear response theory to estimate $U$~\cite{Gironcoli_2005} can lead to an underestimation of Heisenberg exchange interactions, which, in turn, results in underestimated transition temperatures for certain antiferromagnetic (AFM) materials, particularly 3d transition metal oxides (TMOs)~\cite{Mosleh2023}.

TMOs exhibit unique electronic and magnetic properties, making them crucial for applications in batteries, sensors, and catalysts~\cite{song2018transition, xiang2022building, li2017high}. Understanding these properties is essential for advancing new materials and optimizing existing technologies. However, accurate modeling of TMOs presents significant challenges for first-principles methods due to the complex electronic structure of localized $d$ orbitals in transition metals~\cite{tokura2000orbital, rohrbach2003electronic}. As a result, TMOs serve as rigorous test cases for evaluating the strengths and limitations of {\it ab initio}  methods.

While DFT+$U$ offers improvements over standard DFT, it is essential to assess how effectively this approach ---and other {\it ab initio}  methods--- captures the intricate magnetic properties of TMOs. These properties are closely tied to the choice of exchange-correlation approximations. Comparing DFT+$U$ results with more sophisticated methods is a key step toward refining its accuracy. However, modeling bulk TMOs is computationally demanding. A practical alternative is to study simpler systems, such as one-dimensional (1D) chains, which serve as an effective testing ground for evaluating DFT+$U$ against more advanced methods.

Using 1D chains as models in {\it ab initio}  studies is a well-established approach for developing and testing quantum chemistry methods. Hydrogen chains, for example, have been extensively studied in quantum chemistry and materials science~\cite{H_chain1,H_chain2,H_chain3, H_chain4, H_chain5}. In this work, we investigate one-dimensional transition metal oxide chains (1D-TMOs)—VO, CrO, MnO, FeO, CoO, and NiO—as model systems for benchmarking different computational methods. Although 1D-TMOs can be synthesized on surfaces~\cite{MnO2_chain1, MnO2_chain2}, we propose them as ideal models for testing and improving computational techniques. With fewer electrons than their bulk counterparts, 1D-TMOs provide a simplified yet realistic platform for assessing the performance of advanced {\it ab initio}  methods.

Our study systematically evaluates a range of computational methods, including DFT, DFT+$U$, and the highly accurate but computationally expensive coupled-cluster method with single and double excitations (CCSD). We analyze the structural stability, electronic properties, and magnetic ordering of these 1D-TMOs, highlighting the strengths and limitations of each approach. In addition, we examine critical issues such as the influence of pseudopotentials vs full-potential calculations, convergence challenges, and the effects of structural distortions on electronic behavior.

In addition to their computational advantages, one-dimensional TMOs offer a useful platform for comparison with bulk systems. In summary, this paper provides a comprehensive analysis of {\it ab initio}  methods applied to 1D-TMOs, offering insights into the challenges of modeling these systems. Prior studies have examined the magnetic and electronic properties of bulk 3d TMOs using DFT+$U$ and related methods~\cite{bulk1, bulk2, bulk3, bulk4, dzvp-sung}. These works highlight the sensitivity of DFT+$U$ predictions to the choice of $U$ and the complexity of modeling strongly correlated systems. By comparing our 1D results with these bulk benchmarks, we assess which trends are preserved across dimensionalities and which are more sensitive to the structural environment. These comparisons are discussed in Section.~\ref{result} to further evaluate the transferability of {\it ab initio}  methods. The paper is organized as follows: Section,~\ref{sec-method} details the computational methods, Section.~\ref{result} presents our findings, including an analysis of convergence challenges and a comparison of DFT+$U$ with CCSD, and Section.~\ref{conclusion} concludes with a summary of our results.

\section{COMPUTATIONAL METHODS}\label{sec-method}

We use three computational codes for our calculations: Quantum ESPRESSO (QE)~\cite{quantum-espresso}, a plane-wave pseudopotential code; PySCF~\cite{PySCF}, a Python-based quantum chemistry framework; and FHI-aims~\cite{FHI-aims}, an all-electron, full-potential electronic structure code. For density functional theory (DFT) calculations, we employ the Perdew–Burke–Ernzerhof (PBE) functional~\cite{PBE}, a generalized gradient approximation (GGA) for the exchange-correlation energy. In DFT+$U$ calculations, we incorporate Dudarev’s formulation~\cite{Dudarev} as implemented in QE.

To determine the Hubbard $U$ parameter self-consistently for each lattice constant, we employ density functional perturbation theory (DFPT)~\cite{DFPT_RMP,Timrov_2018, Timrov_2021}, which enables $U$ to be calculated based on the linear response method~\cite{Gironcoli_2005} as implemented in QE. For CCSD, we use PySCF.

\begin{figure}
\includegraphics[scale=0.20]{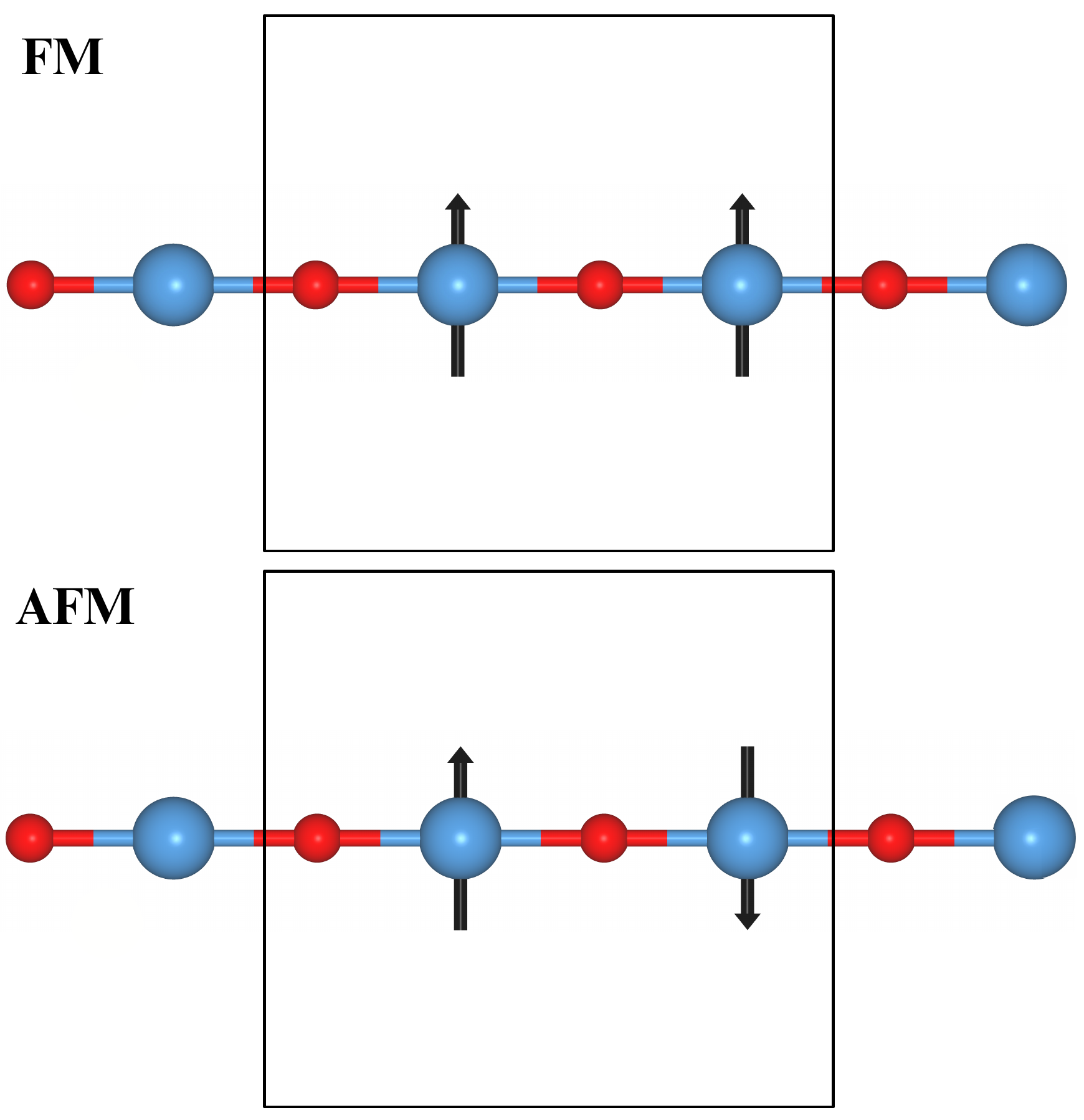} 
\caption{\label{shematic}
The crystal structure of a one-dimensional (1D) transition-metal monoxide (TMO) chain, oriented along the $x$-axis, is shown. The transition metal (TM = V, Cr, Mn, Fe, Co, Ni) and oxygen atoms are represented by blue and red circles, respectively. The arrows indicate the magnetic moments of the transition metal atoms, illustrating two magnetic configurations: FM and AFM states.}
\end{figure}

Our study focuses on first-row transition metal mono-oxides (TMOs, with TM = V, Cr, Mn, Fe, Co, and Ni) arranged in a one-dimensional (1D) chain structure along the $x$-direction (Fig.~\ref{shematic}). Each chain is investigated in two magnetic configurations: ferromagnetic (FM) and antiferromagnetic (AFM). For AFM states, we use a minimal unit cell containing two formula units (four atoms) to account for magnetic ordering, and for the FM states, we adopt the same geometry unless explicitly stated. To minimize interactions between periodic images, we introduce a vacuum thickness of 30 atomic units (a.u.), and the Brillouin zone is sampled using a $4\times1\times1$ $k$-point mesh.

In QE, GBRV ultra-soft pseudopotentials are used~\cite{GBRV} with a kinetic energy cutoff of 60 Ry for the plane-wave expansion. For all-electron calculations in FHI-aims, we select a tight-tier2 basis set. In PySCF calculations, we employ the Goedecker–Teter–Hutter (GTH) pseudopotential with the corresponding GTH-DZVP-MOLOPT-SR basis set (abbreviated as DZVP)~\cite{basis-gth,dzvp-sung}, obtained from the CP2K library~\cite{cp2k}, to represent the valence electrons. Unrestricted CCSD calculations are carried out with initial reference states generated from unrestricted Hartree-Fock (UHF).

To compare the computational methods and determine the most stable magnetic state, we calculate the energy difference, $\Delta E = E_{\mathrm{AFM}} - E_{\mathrm{FM}}$, between the AFM and FM phases at their optimal geometries, where $E_{\mathrm{AFM}}$ and $E_{\mathrm{FM}}$ denote the energies of the AFM and FM states, respectively.

\section{RESULTS}\label{result}

\subsection{Instability and convergence issues}
\begin{figure}[!htb]
    \centering
	\includegraphics[width=8cm ,scale=1]{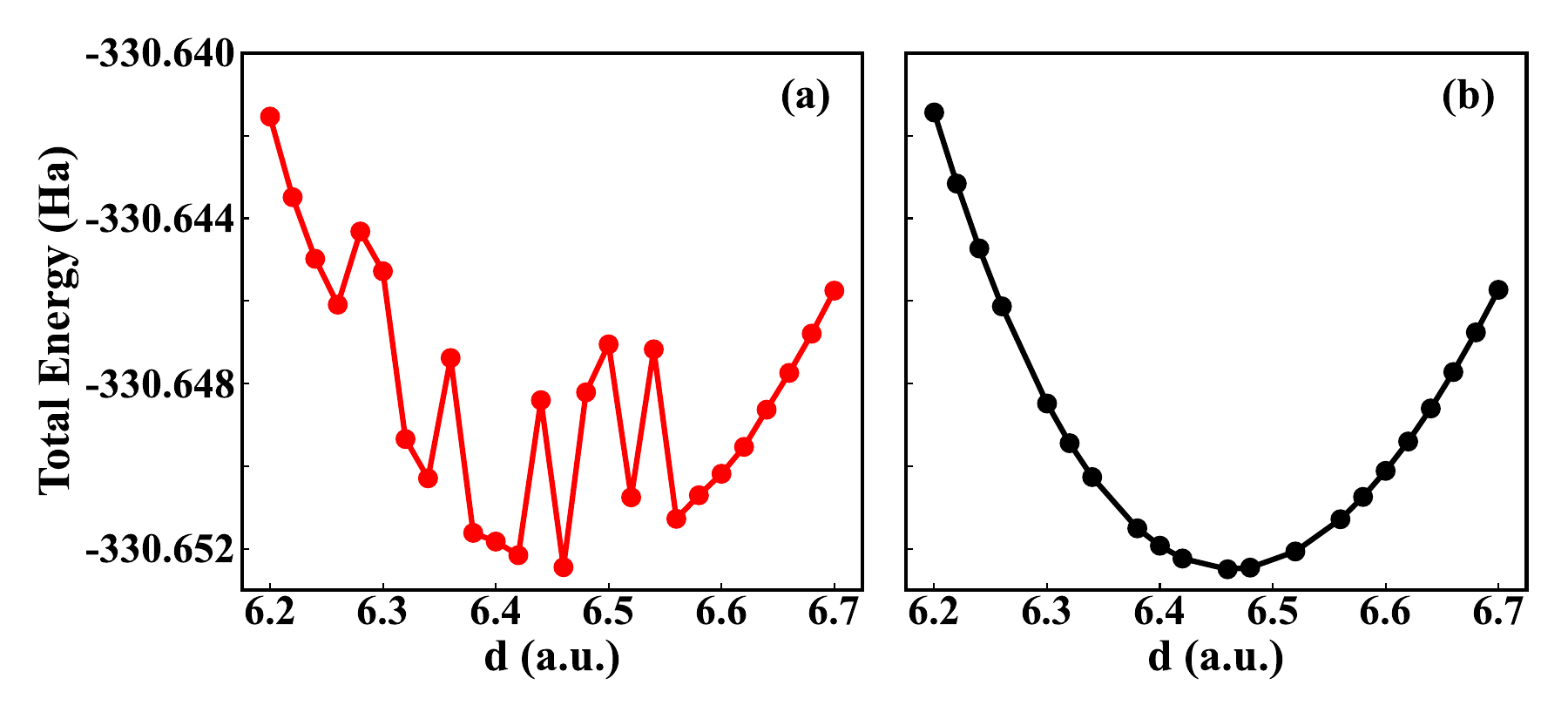} 
	\caption{\label{instability}Variation in the total energy vs lattice constant (\textbf{d}) for CoO using the PBE method calculated by Quantum EEPRESSO, in the AFM phase :(a) Initial calculations showing instability in the wavefunction. (b) Final calculations after applying our trick to address the instability, demonstrating improved convergence. } 
\end{figure}

With the exception of the MnO chain, which shows stable convergence, all PBE and DFT+$U$ calculations---regardless of the DFT code used (i.e., PySCF, QE, and FHI-aims)---face significant wavefunction instability issues, often causing the self-consistent field (SCF) calculations to converge to an excited state instead of the ground state. Detecting such instabilities only requires comparing the total energy of the 1D-TMOs at different lattice parameters. Fig.~\ref{instability}(a) illustrates the variation in total energy with respect to the lattice parameter (TM–TM distance) for the CoO chain in the AFM state. Although the calculations converge successfully for all lattice parameters, some converge to unstable states, resulting in a zigzag energy plot rather than the expected smooth curve. This behavior indicates that some SCF calculations fail to reach a ground-state solution.

\begin{figure}[!htb]
\centering
\includegraphics[width=9 cm ,scale=0.5]{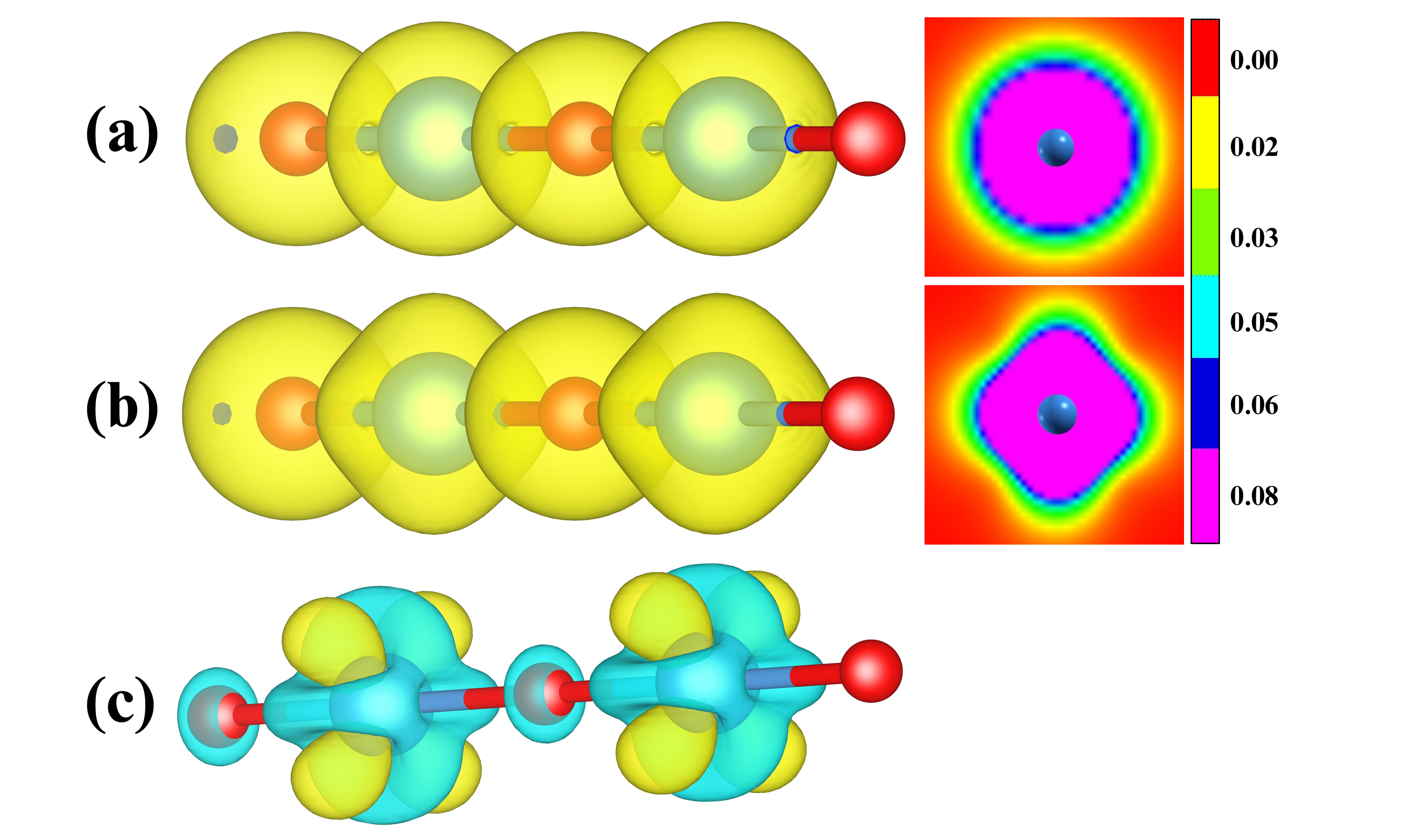}
\caption{\label{charge_density_profile}
(Colour online) The isosurface of the spin density for CoO in the FM state: (a) The spin density is isotropic for the unstable configuration. (b) An anisotropic spin density is observed for the stable configuration. (c) Shows the difference in spin density (subtracted). A slice of the spin density centered on the Co atom in the (011) surface is shown in the right panel.} 
\end{figure}

Projected density of states (PDOS) calculations reveal differences in $d$-orbital occupations between unstable and stable points, suggesting that these instabilities may arise from the limitations of {\it ab initio}  methods in determining optimal electronic $d$-orbital occupations.
Fig.~\ref{charge_density_profile} presents the spin-down charge density profile of CoO in a (011) surface cut, showing significant differences between stable and unstable configurations.
This suggests that the $d$-orbital degrees of freedom in one-dimensional transition metal oxides (1D-TMOs) can contribute to instability in {\it ab initio}  calculations.

For a given SCF calculation, PySCF performs stability analysis to detect and correct wavefunction instabilities by introducing small perturbations and checking whether the perturbed state leads to a lower energy~\cite{Seeger1977}. However, our study reveals that this process often becomes trapped in quasi-stable states, making it unreliable for consistently reaching the ground state.

To overcome this issue, we adopt a straightforward yet effective stabilization procedure. We first examine the total energy profile [Fig.~\ref{instability}(a)] along with the magnetic moments of the transition metal atoms. When the energy profile shows irregularities—such as zigzag patterns or unexpected jumps—we identify lattice points that exhibit physically consistent solutions: smooth energy variation, lower total energy, symmetric magnetic moments on equivalent atoms, and correct magnetic ordering (e.g., AFM or FM as expected). These configurations are considered stable and used as reference points.

For unstable points (i.e., those with irregular energies or asymmetric local moments), we restart the SCF using the converged wavefunction from a nearby stable configuration as the initial guess. This targeted initialization helps the SCF escape from local excited-state minima and converge toward the correct ground-state solution. This approach is conceptually similar to Direct Inversion in the Iterative Subspace (DIIS) because it uses previously obtained stable solutions to guide convergence. Using this method, we successfully obtain smooth and physically meaningful total energy curves, as shown in Fig.~\ref{instability}(b). We apply this approach consistently across all DFT and HF calculations.

\subsection{Structural properties}
\begin{figure*}[t]
\centering
\includegraphics[width=18.0cm, scale=1.0]{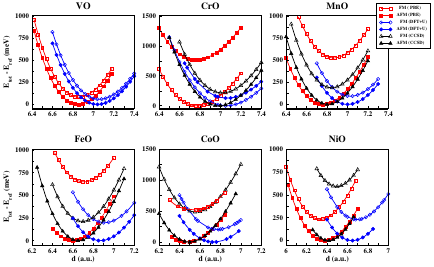}
\caption{Variation of total energy with respect to the lattice constant (\textbf{d}). The reference energy ($E_{\mathrm{ref}}$) corresponds to the minimum energy of the stable phase. The calculations were performed using the PBE and DFT+$U$ methods with Quantum ESPRESSO and the CCSD method with PySCF for the FM and AFM phases of 1D-TMO chains.}
\label{eng_pbe}
\end{figure*}

To study the electronic structure and magnetic properties of 1D-TMOs, we first determine their most stable atomic configurations. We perform a series of calculations for different lattice constants, considering both FM and AFM states. Fig.~\ref{eng_pbe} illustrates how the total energy varies with the lattice parameter for each chain, computed using different {\it ab initio}  methods.

Our results indicate that the AFM state is the most stable configuration for all chains, except CrO, when using both PBE and DFT+$U$. 
We obtain the optimized lattice constants, corresponding to the minimum energy values, by fitting the total energy data to the Birch–Murnaghan equation of state~\cite{Murnaghan,Birch}. These optimized lattice constants are then used to determine other properties, such as the magnetic moments of the 1D-TMOs.

To calculate the magnetization, we analyze the PDOS of these systems. Table.~\ref{combined_table} summarizes the ground-state lattice constants, magnetization, and energy parameter ($\Delta E$) for each chain. The table clearly shows that the FM state generally results in slightly larger lattice constants and higher magnetic moments. However, within the PBE approach, the NiO chain is an exception, where the AFM state shows a larger lattice constant and higher magnetic moment compared to the FM state.

In addition, the magnetization of NiO is lower than expected within PBE, which can be attributed to its electronic structure near the Fermi level, where the orbitals are half-filled. This finding highlights the necessity of applying a Hubbard $U$ correction through DFT+$U$ calculations.

\begin{table*}
\caption{The calculated lattice parameters (\textbf{d}) and magnetic moments per ion in the AFM and FM phases, as well as the energy differences ($\Delta E$) between the two magnetic phases using the PBE method in QE, FHI-aims, and PySCF; the DFT+$U$ method in QE; and the CCSD method in PySCF. The absence of results for VO in PBE (PySCF) is due to the need for basis sets larger than TZVP to achieve convergence. For VO, we use fixed $U$ values of 4.8 eV, while for the other compounds, we employ a self-consistent $U$ parameter determined using DFPT.}
\label{combined_table}
\begin{ruledtabular}
\begin{tabular}{@{}lcccccc@{}}
 & \textbf{}            & \multicolumn{2}{c}{\textbf{\begin{tabular}[c]{@{}c@{}}Lattice constant\\ a.u\end{tabular}}} & \multicolumn{2}{c}{\textbf{\begin{tabular}[c]{@{}c@{}}Magnetic/ion\\$\mu_B$\end{tabular}}} & \textbf{\begin{tabular}[c]{@{}c@{}}$\Delta E$\\ meV\end{tabular}} \\
\hline
 & \multicolumn{1}{l}{} & \multicolumn{1}{l}{\textbf{AFM}}  & \textbf{FM}  & \multicolumn{1}{l}{\textbf{AFM}}  & \textbf{FM}  & \multicolumn{1}{l}{}      \\  [1ex]

\textbf{VO}  & PBE (QE)       & 6.85  &6.86         & 3.09  &3.15         & -82      \\
             & PBE (FHI-aims) & 6.85  &6.86         & 3.06  &3.16         & -116     \\
             & PBE (PySCF)    & ---  & ---          & ---   & ---         & ---      \\
             & DFT+$U$ (QE)   & 7.04  &7.06         & 3.16  &3.23         & -58      \\
\hline
\textbf{CrO} & PBE (QE)       & 6.76  &6.77         & 3.78  &4.11         & +764     \\
             & PBE (FHI-aims) & 6.75  &6.76         & 3.73  &4.20         & +764     \\
             & PBE (PySCF)    & 6.75  &6.74         & 3.71  &4.00         & +854     \\
             & DFT+$U$ (QE)   & 7.06  &7.06         & 4.11  &4.45         & +128     \\
             & CCSD (PySCF)   & 6.98  &7.02         & ---   & ---         & -215     \\
             
\hline
\textbf{MnO} & PBE (QE)       & 6.76  &6.87         & 4.55  &4.80         & -526     \\
             & PBE (FHI-aims) & 6.75  &6.86         & 4.44  &4.82         & -541     \\
             & PBE (PySCF)    & 6.74  &6.83         & 4.63  &4.79         & -467     \\
             & DFT+$U$ (QE)   & 7.00  &7.03         & 4.89  &4.97         & -93      \\
             & CCSD (PySCF)   & 6.89  &6.85         & ---   & ---         & -188     \\
             
\hline
\textbf{FeO} & PBE (QE)       & 6.60  &6.71         & 3.44  &3.72         & -671     \\
             & PBE (FHI-aims) & 6.57  &6.66         & 3.20  &3.67         & -506     \\
             & PBE (PySCF)    & 6.60  &6.72         & 3.44  &3.65         & -613     \\
             & DFT+$U$ (QE)   & 6.87  &6.91         & 3.88  &3.93         & -200     \\ 
             & CCSD (PySCF)   & 6.53  &6.70         & ---   & ---         & -211     \\

\hline
\textbf{CoO} & PBE (QE)       & 6.47  &6.54         & 2.21  &2.44         & -514     \\
             & PBE (FHI-aims) & 6.32  &6.44         & 1.86  &1.46         & -490     \\
             & PBE (PySCF)    & 6.46  &6.56         & 2.26  &2.54         & -239     \\
             & DFT+$U$ (QE)   & 6.73  &6.78         & 2.70  &2.79         & -206     \\
             & CCSD (PySCF)   & 6.50  &6.56         & ---   & ---         & -483     \\

\hline
\textbf{NiO} & PBE (QE)       & 6.38  &6.33         & 0.99  &0.41         & -250     \\
             & PBE (FHI-aims) & 6.36  &6.32         & 0.79  &0.43         & -300     \\
             & PBE (PySCF)    & 6.36  &6.45         & 1.22  &1.45         & -366     \\
             & DFT+$U$ (QE)   & 6.65  &6.69         & 1.66  &1.75         & -230     \\
             & CCSD (PySCF)   & 6.41  &6.49         & ---   & ---         & -596     \\ [1ex]             
         
\end{tabular}
\end{ruledtabular}
\end{table*}

In the next step, to validate the pseudopotentials, we conducted a comparison between results obtained using the full-potential FHI-aims and pseudopotentials from QE, employing the GGA exchange-correlation functional. The calculated parameters are presented in Table.~\ref{combined_table}. The lattice parameters calculated using QE and FHI-aims show good agreement, with differences generally within 0.01–0.02 atomic units (a.u.), indicating the reliability of the present pseudopotentials approach in predicting the lattice constants. The Mulliken population analysis to estimate atomic spins also shows consistency between the two methods, though some differences are notable. For example, the magnetic moments of VO, CrO and MnO show small variations. The differences are more pronounced in compounds such as FeO (AFM), CoO (FM), and NiO (AFM). For the AFM state of FeO, the magnetization difference calculated by the two codes is 0.24 $\mu_B$; for the FM state of CoO it is 0.98 $\mu_B$, and for the AFM state of NiO the difference is about 0.2 $\mu_B$.

The energy differences between the AFM and FM phases ($\Delta E$) are generally consistent between the two methods, although there are some discrepancies. For VO, CrO, MnO, CoO, and NiO, the difference in $\Delta E$ between the methods is less than or equal to 0.05~eV in all cases. Only for FeO does this difference exceed 0.16~eV.

In summary, the GBRV pseudopotentials in QE generally provide results in close agreement with the full-potential FHI-aims calculations for lattice parameters, magnetic moments, and energy differences, validating their use in predicting the properties of these transition metal monoxide chains. The slight variations observed underscore the nuanced differences between pseudopotentials and all-electron methods, emphasizing the importance of method selection based on the specific requirements of a study. 

Although the results of QE and FHI-aims for NiO are quite similar, there are notable differences. Unlike other chains, the AFM state leads to larger lattice constants and higher magnetic moments. Moreover, even full-potential calculations fail to accurately capture the magnetization of the FM state. However, using DFT+$U$ yields more reasonable results for NiO.

The PySCF-PBE results show a strong dependence on the choice of basis sets for VO. Specifically, $\Delta E$ is positive with SVP and DZVP but approaches zero with TZVP. Due to this significant basis set dependency, we exclude VO from further CCSD investigations.

For other cases, DZVP and TZVP yield similar results. However, discrepancies persist between PySCF-PBE calculations for CoO and NiO compared to FHI-aims and QE. This highlights the need for further investigation into basis sets and pseudopotentials for TMOs. Nevertheless, since larger basis sets make CCSD calculations impractical for our system, we limit our calculations to the DZVP basis set.

Our findings suggest that 1D-TMOs provide a useful testing ground for benchmarking DFT codes with different computational approaches.

\subsection{Electronic properties}

\begin{figure*}
    \centering
	\includegraphics[width=17.0 cm ,scale=0.91]{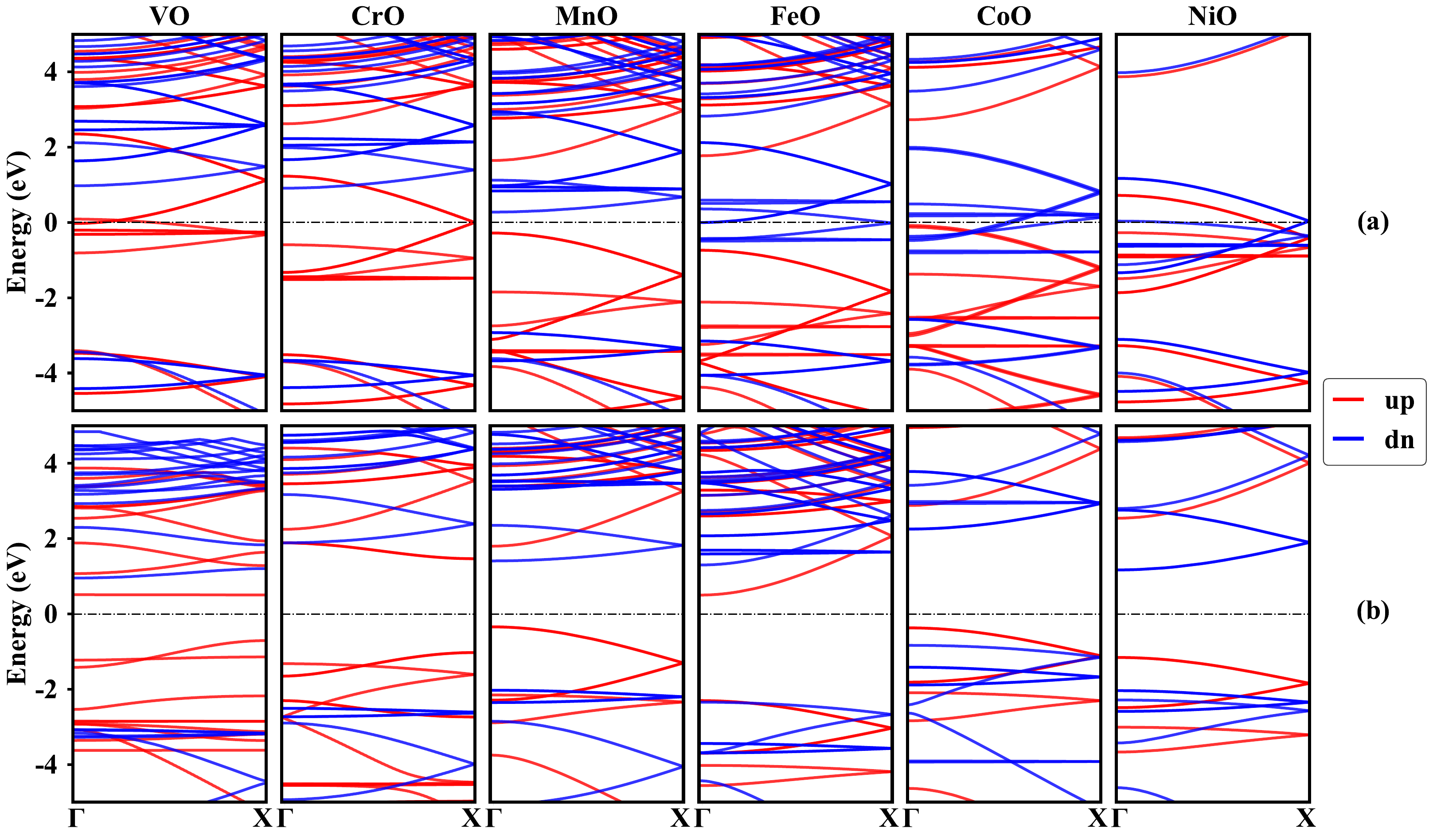} 
	\caption{Calculated band structure of all considered systems in the FM magnetic state. The Fermi level $E_{F}$ is chosen as the zero of the energy scale. Effective values of $U$ as determined using the QE package in Table.~\ref{hubb_U} are used in the DFT+$U$ calculation. Top panels show the PBE band structures and the bottom panels show the DFT+$U$ band structures. For VO, the $U$ parameteris  set to 4.8 eV~\cite{Uparam}.
    The red and blue lines represent the spin-up (up) and spin-down (dn) band structures, respectively.
    }\label{band_pbe-fm}
\end{figure*}

The study of transition metal oxides (TMOs) is challenging due to the presence of localized 3$d$ orbitals in transition metal atoms. The PBE approximation suffers from self-interaction errors, which prevent it from accurately describing these localized orbitals. This limitation is evident in the PBE band structures of the FM state of 1D-TMOs, where only MnO exhibits a band gap (see Fig.~\ref{band_pbe-fm}). To correctly open the band gap, PBE must be supplemented with correction methods such as DFT+$U$. The converged $U$ values, calculated at the optimal lattice parameter, are presented in Table.~\ref{hubb_U}. 

The calculated Hubbard $U$ values show only a small difference (\~0.1–0.2 eV) between the AFM and FM states, without exhibiting a specific trend. Moreover, there is no clear relationship between the $d$-orbital filling and the $U$ value in either magnetic state. This suggests that, although the Hubbard $U$ is slightly influenced by magnetic ordering, it primarily depends on the electronic environment and the atomic species.

For VO, we encountered serious convergence problems when trying to compute the Hubbard $U$ value self-consistently using DFPT. The SCF calculations were unstable across different lattice parameters and resulted in large, unphysical fluctuations in total energy. Because of this, we were unable to obtain a reliable $U$ value and instead used a fixed value of 4.8 eV for V~\cite{Uparam}. This approach ensured numerical stability and produced results that are consistent with the general behavior observed in other TMOs.

 \begin{figure*}
    \centering
	\includegraphics[width=17.0 cm ,scale=0.9]{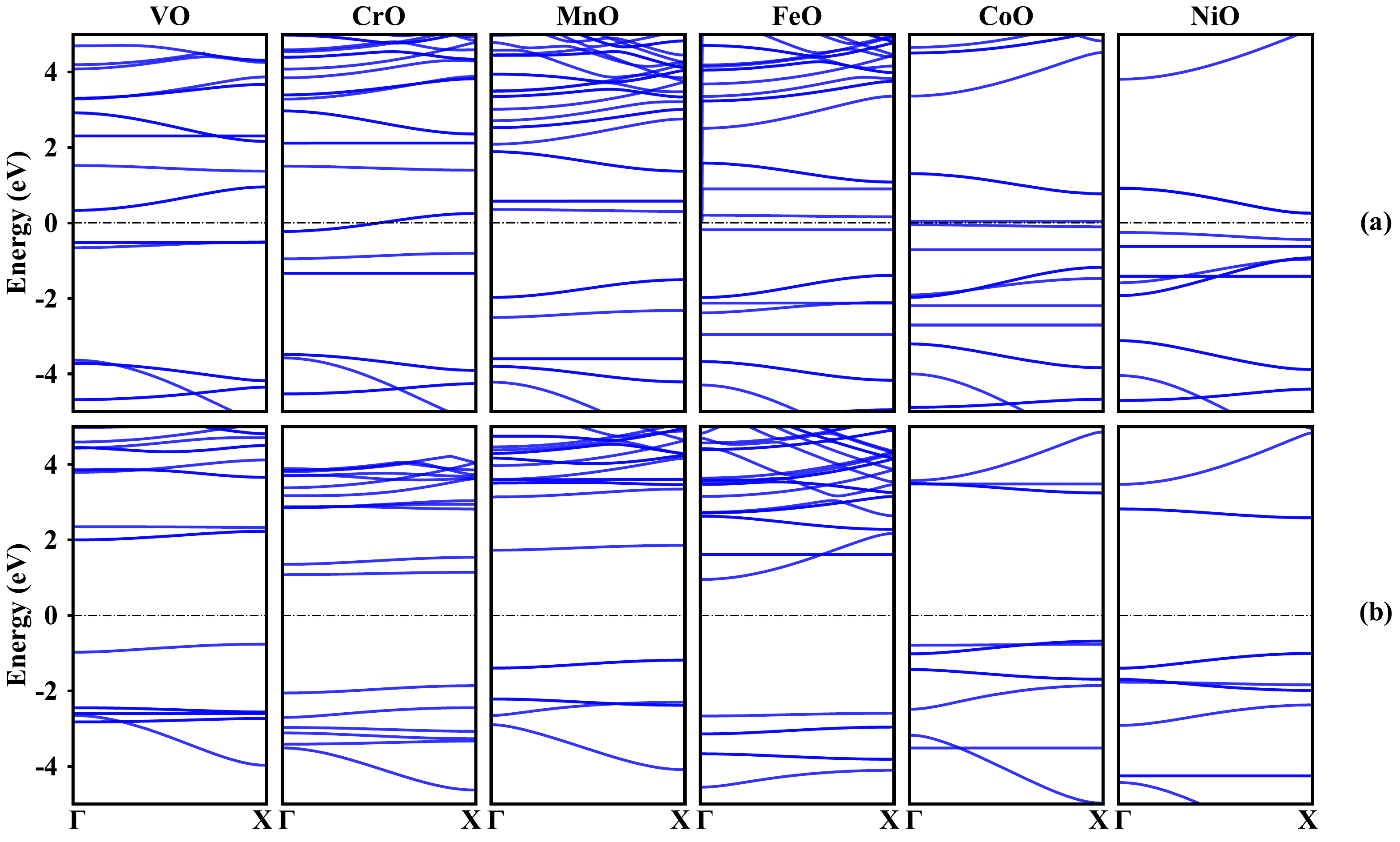} 
	\caption{Calculated band structure of all considered systems in the AFM magnetic state. The Fermi level $E_{F}$ is chosen as the zero of the energy scale. Effective values of $U$ as determined using the QE package in Table.~\ref{hubb_U} are used in the DFT+$U$ calculation. Top panels show the PBE band structures and the bottom panels show the DFT+$U$ band structures. For VO, the $U$ parameter is set to 4.8 eV~\cite{Uparam}.}\label{band_pbe-af}
\end{figure*}

\begin{table}
\caption{Converged value of the effective Hubbard $U$ parameter (in eV) for the transition metal(TM) 3d orbitals for FM and AFM states.} \label{hubb_U}
\begin{ruledtabular}
\begin{tabular}{@{}cccccc@{}}
\multicolumn{1}{l}{} Magnetic state& CrO & MnO & FeO & CoO & NiO \\[0.1ex]
\hline
FM   & 6.57  & 4.39   & 7.46  & 5.97  &7.17 \\
AFM  & 6.47  & 4.58   & 7.28  & 6.09  &7.30 \\[1ex]
\end{tabular}
\end{ruledtabular}
\end{table}

Fig.~\ref{band_pbe-fm} and ~\ref{band_pbe-af} show the band structures of the 1D-TMOs computed using the PBE (top panel) and DFT+$U$(bottom panel) methods for AFM and FM magnetic states, respectively.

For the FM states of VO, FeO, and CoO, the PBE method predicts a half-metallic character, while it predicts a fully metallic state for NiO and a semi-metallic state for CrO. Applying DFT+$U$ results in an insulating state for all 1D-TMOs. 
For NiO, while our PBE calculations predict a metallic ferromagnetic state, applying DFT+$U$ opens a band gap. In a study on an ideal infinite NiO chain using a hybrid functional~\cite{NiO-chain}, a metastable half-metallic solution was obtained for the ferromagnetic ordering, underscoring the importance of carefully identifying the global minimum in the electronic structure of 1D-TMOs.

Among the TMOs we studied, MnO is the only system that exhibited stable convergence across all PBE and DFT+$U$ calculations, regardless of the code used, and showed no signs of spurious local minima or metastable solutions. In addition, both PBE and DFT+$U$ consistently predict an insulating ground state for MnO in our 1D chain calculations. This is consistent with bulk studies, which also report an insulating character at the PBE level—typically yielding a gap of ~1.1eV—and larger values (~3.5eV) from more accurate methods such as CCSD[~\cite{dzvp-sung, MnO-insulate2}]. This combination of numerical stability and robust insulating behavior makes MnO a particularly suitable candidate for benchmarking correlated electronic structure methods across different dimensionalities.

In the AFM state (Fig.~\ref{band_pbe-af}), both the PBE and DFT+$U$ methods predict an insulating ground state for VO, MnO, FeO, and NiO. 
The $U$ correction pushes the occupied bands down and opens the energy gap. In the case of CrO and CoO, PBE fails to open a gap, predicting a metallic ground state. However, the $U$ correction succeeds in predicting insulating behavior.

For CrO in the FM state, a straightforward application of DFT+$U$ does not produce an insulating solution. To assess whether the FM state is inherently semi-metallic or if this result arises from the system being trapped in a local minimum, it is important to test various initializations of the $d$-orbital density matrix within DFT+$U$\cite{OMC1, OMC2}. A simpler and more effective alternative is to randomly initialize the magnetic moments on the oxygen atoms~\cite{SMC}. Using this approach, we find that the FM state of CrO can indeed transition to an insulating state.

\begin{figure}[!htb]
\centering
\includegraphics[width=8.5 cm ,scale=1]{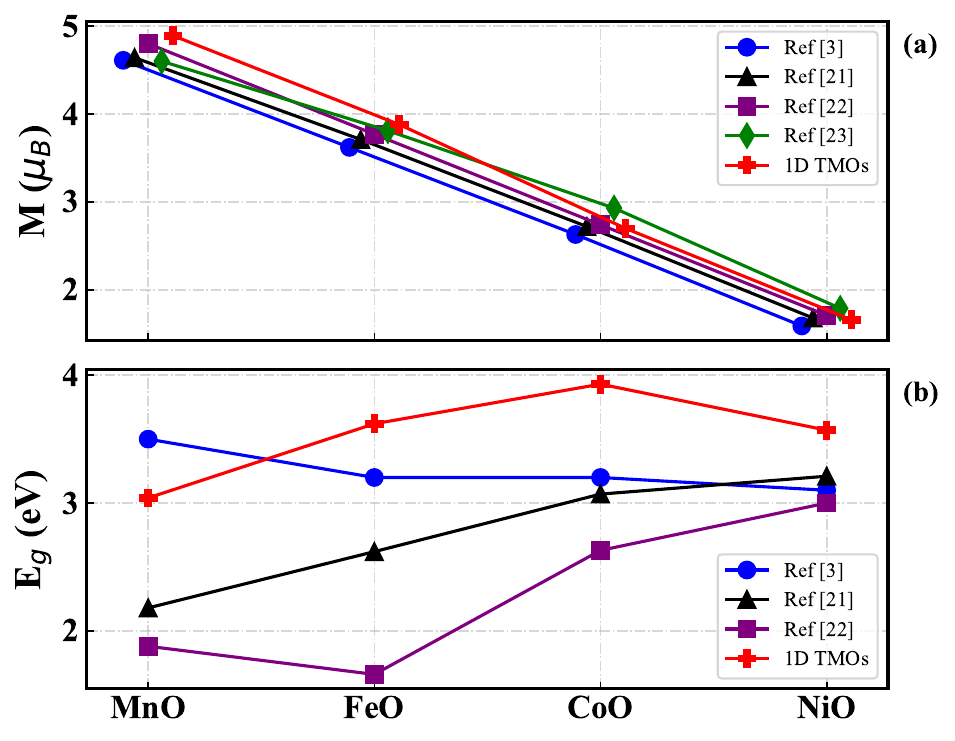}
\caption{Trend of (a) local magnetic moments (M) in $\mu_B$ and (b) band gaps (Eg) in eV for selected transition metal monoxides (MnO, FeO, CoO, NiO) as obtained for 1D-TMOs and from reported DFT+$U$ studies on bulk systems. In panel (a), the data points are slightly shifted along the $x$-axis for clarity.}
\label{gap-trend}
\end{figure}

Fig.~\ref{gap-trend}(a) compares our calculated magnetic moments for MnO, FeO, CoO, and NiO with those reported in previous DFT+$U$ studies of bulk systems[~\cite{DFTU1, bulk2, bulk3, bulk4}]. Due to low coordination, magnetic moments in 1D-TMOs are generally expected to be larger than in bulk. While our results qualitatively follow bulk trends, the magnitudes are not consistently higher. These deviations are primarily due to differences in computational methodology. In particular, definitions of local moments vary across studies—e.g., via atomic sphere integration or orbital projections. Here, we extract magnetic moments from the PDOS.

Fig.~\ref{gap-trend}(b) shows that the calculated band gaps in the 1D-TMO chains are systematically larger than those in bulk[~\cite{DFTU1, bulk2, bulk3}], consistent with enhanced electron localization in reduced dimensions. For all four compounds, the 1D gaps exceed bulk DFT+$U$ values by 0.3–1.2eV. While this reflects stronger correlation effects in 1D, the precise values remain sensitive to the choice of $U$, and the exchange-correlation functional.

Overall, the observed trends in both magnetic moments and band gaps indicate stronger localization and enhanced correlation effects in the 1D models compared to their bulk counterparts. This systematic behavior underscores the utility of 1D-TMOs as simplified yet physically meaningful platforms for benchmarking electronic structure methods in strongly correlated systems.

\subsection{Comparison between DFT+\texorpdfstring{$U$} and CCSD}

CCSD is known for its high accuracy in electronic structure predictions, but it comes with a significantly higher computational expense compared to other methods, such as DFT+$U$, due to its steep scaling with system size. Due to having fewer of electrons in 1D-TMOs, they can be ideal test cases to compare DFT+$U$ results with CCSD.

The total energy vs lattice parameter for 1D-TMOs is calculated using CCSD (Fig.~\ref{eng_pbe}). The calculated energy differences, $\Delta E$, (in meV) for CrO, MnO, FeO, CoO, and NiO chains using DFT+$U$ and CCSD methods (starting from UHF orbitals) are presented in Table.~\ref{combined_table}. Notably, the CCSD method generally predicts higher energy differences compared to DFT+$U$. For CrO, MnO, CoO, and NiO, the CCSD results show significantly higher values for $\Delta E$ with respect to the DFT+$U$. In contrast, FeO demonstrates closer agreement between the two approaches, with only a slight deviation in the energy difference. 

The largest discrepancies are observed for NiO and CrO. For NiO, DFT+$U$ predicts an energy difference of $\Delta E = -230$ meV, while CCSD yields a much lower value of $\Delta E = -596$ meV, highlighting the challenges of accurately capturing electron correlation in this material. For CrO, the discrepancy is even more striking: DFT+$U$ predicts the ferromagnetic state as the ground state with $\Delta E = +128$ meV, whereas CCSD identifies the antiferromagnetic state as the true ground state with $\Delta E = -215$ meV.

The discrepancy between PBE, DFT+$U$ and CCSD in predicting the ground state of CrO may partly stem from the use of unrestricted Hartree-Fock (UHF) as the reference for CCSD calculations, as UHF itself favors the antiferromagnetic state as the ground state of CrO. Since CCSD builds on UHF reference wavefunctions, this preference may influence the final result. We find that both FHI-aims and PySCF predict the AFM ground state for CrO at the UHF level, suggesting that this outcome is likely robust with respect to the choice of basis set.

These results for MnO, FeO, CoO, and NiO support our previous findings~\cite{Mosleh2023}, indicating that the linear response method can, in some cases, overestimate the Hubbard $U$ value.

This overestimation can lead to an underestimation of the energy difference between ferromagnetic and antiferromagnetic states, ultimately resulting in a lower predicted transition temperature.

The equilibrium lattice constants of the CCSD closely match the PBE results, whereas DFT+$U$ predicts lattice constants that are 3\%–4\% larger. This discrepancy highlights the impact of the Hubbard $U$ correction when combined with a GGA functional like PBE, which tends to overestimate lattice constants—a well-known issue in DFT+$U$ calculations (see Fig.~\ref{eng_pbe} and Table.~\ref{combined_table}).

It is worth noting that CCSD is more sensitive to the choice of basis set functions compared to DFT+$U$. However, due to limited computational resources, we are unable to evaluate the impact of the basis set on CCSD results. We hope that future studies will achieve more accurate CCSD calculations using larger basis sets.

\section{CONCLUSION}\label{conclusion}

In summary, we investigated the structural, electronic, and magnetic properties of one-dimensional transition-metal oxide chains using a range of ab initio methods. We began by addressing convergence and instability issues that arose in our DFT calculations, particularly for CoO, and NiO, where wavefunction instability resulted in convergence to incorrect states rather than ground states. Using stable wavefunctions at some lattice parameters as the starting point, we were able to overcome these challenges and achieve stable ground-state configurations. For ground state properties, we compared results from pseudopotential (QE) and full-potential (FHI-aims) methods. The results showed general agreement in lattice constants and energy differences.

We also examined the band structures of these chains using PBE and DFT+$U$ methods. For the FM state, PBE predicted half-metallic or metallic ground states in most cases, whereas the DFT+$U$ approach successfully opened band gaps. For the AFM state, PBE predicts an insulating state for VO, MnO, FeO, and NiO, but for CrO and CoO, there is electronic contribution at the Fermi level. However, using DFT+$U$, we were able to open a band gap for both CoO and CrO.

Finally, a comparison between DFT+\(U\) and the more accurate, yet computationally demanding, CCSD method revealed significant discrepancies, particularly for NiO and CrO. In the case of CrO, we encountered a notable contradiction: CCSD predicts an AFM ground state, whereas DFT+\(U\) favors an FM ground state. This makes the CrO chain a compelling candidate for benchmarking and testing the accuracy of various {\it ab initio}  methods.

For all cases, the CCSD method predicts significantly larger energy differences between AFM and FM states compared to DFT+$U$, suggesting that the $U$ parameters obtained through the linear response theory may be overestimated for these systems. However, for a more reliable comparison, CCSD calculations should be performed using larger basis sets. 
Comparison with bulk TMO data reveals that 1D models capture key trends, including the decrease in magnetic moment and the increase in band gap with $d$-orbital filling. Magnetic moments generally follow bulk behavior, while band gaps are consistently larger in 1D systems due to enhanced electron localization from reduced dimensionality. These findings support 1D-TMOs as simplified yet informative benchmarks for correlated electronic structure methods.
Overall, this comprehensive benchmarking highlights the strengths and limitations of each method, providing valuable insights for future research on low-dimensional transition-metal oxides, which serve as a challenging model for {\it ab initio}  calculations.

\section*{DATA AVAILABILITY}
The data that support the findings of this study are available upon reasonable request from the authors.

\section*{ACKNOWLEDGMENTS}
This work was supported by the Vice-Chancellor for Research Affairs of Isfahan University of Technology (IUT). 
S.d.G.’s work was supported by the European Commission through the MAX Center of Excellence for Supercomputing Applications (Grant Nos. 10109337 and 824143) and by the Italian MUR, through the Italian National Center for HPC, Big Data, and Quantum Computing (ICSC, Grant No. CN00000013). Computational resources were provided by CINECA.

\clearpage
\appendix
\section*{APPENDIX: HUBBARD \texorpdfstring{$U$}{U}  PARAMETERS COMPUTED FROM DFPT}

In this appendix, we provide the computed Hubbard $U$ parameters obtained via density-functional perturbation theory (DFPT) for each transition metal oxide (TMO) chain considered in this work. These data correspond to the values used in all DFT+$U$ total energy and electronic structure calculations presented in the main text.

As explained in Sec.~\ref{sec-method} in our DFT+$U$ calculations, the Hubbard $U$ parameter was not treated as an empirical input but instead was computed self-consistently using the linear response approach within DFPT, following the method proposed by Cococcioni and de Gironcoli.~\cite{Gironcoli_2005}, This technique allows for a first-principles determination of $U$ by evaluating the response of the localized $d$-orbitals to a perturbing potential within the same theoretical framework as the DFT functional.

\begin{figure}[ht]
  \centering
  \includegraphics[width=0.95\linewidth]{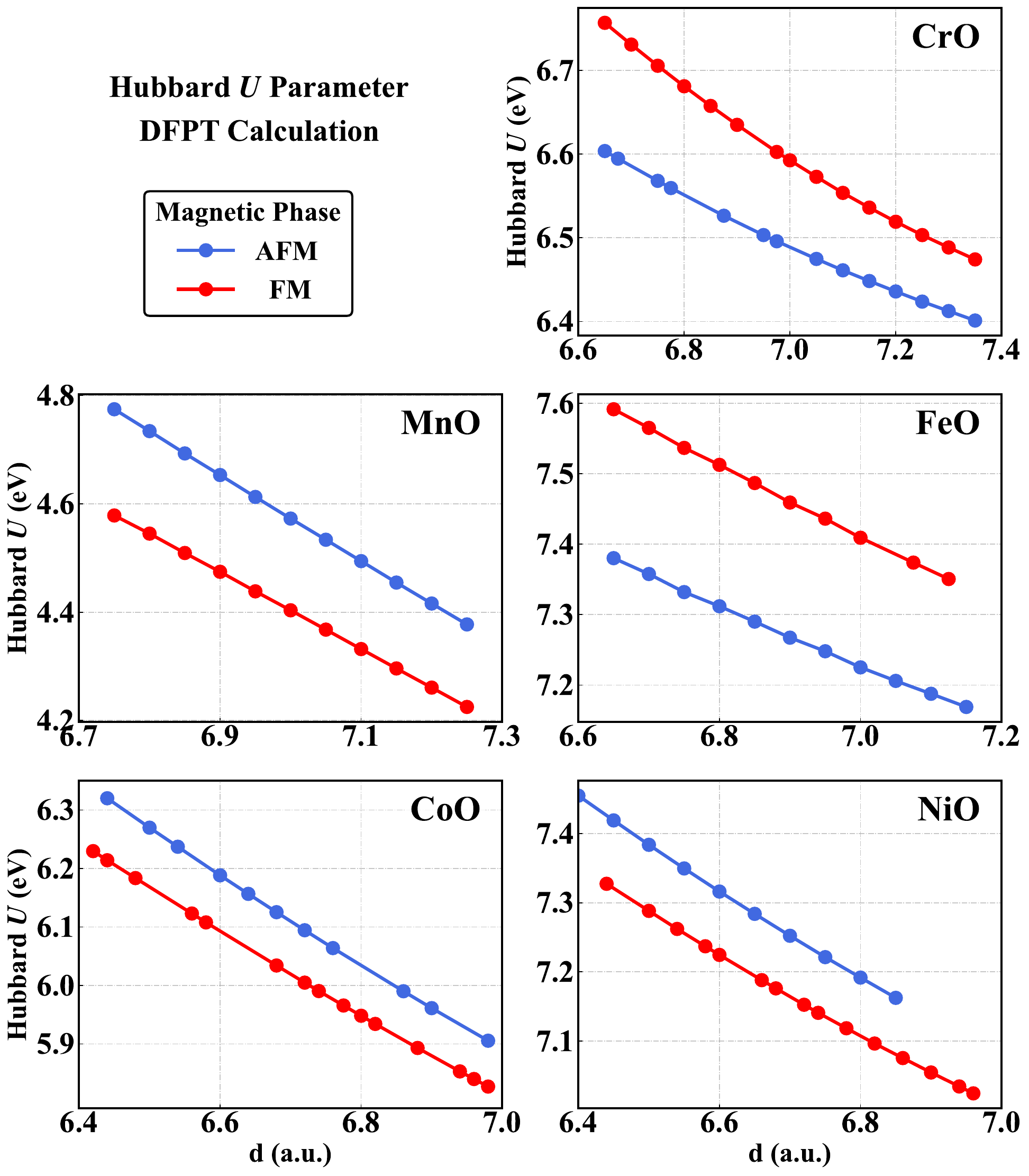}
  \caption{Hubbard $U$ values calculated via DFPT as a function of lattice constant(d) for each TMO chain in both FM and AFM magnetic configurations.}
  \label{fig:DFPT-U}
\end{figure}

For each material (CrO, MnO, FeO, CoO, and NiO), the $U$ parameter was computed as a function of the lattice constant separately for the FM and AFM phases. These values were used consistently in our PBE+$U$ calculations for total energies and electronic structures. Fig.~\ref{fig:DFPT-U} summarizes the computed Hubbard $U$ values across the lattice scans for all systems.

As shown, the $U$ parameter generally decreases with increasing lattice constant, likely due to changes in TM–O hybridization that enhance electronic screening in the expanded structures. In addition, the observed differences between FM and AFM phases indicate a non-negligible dependence of $U$ on the magnetic configuration. These variations are important when computing relative phase stability, as they influence the total energy landscape in DFT+$U$.

\clearpage
\bibliography{reference}

\end{document}